\documentclass[12pt]{article}
\usepackage{epsf}
\usepackage{amsmath,amssymb}
\usepackage{pdfpages}
\usepackage{graphicx}
\usepackage{comment}
\usepackage{tikz}
\usetikzlibrary{positioning}
\usepackage{physics}

\newcommand{\Ldet}{Z}
\newcommand{\lpath}{L}

\begin{document}

\newcommand{\rem}[1]{{$\spadesuit$\bf #1$\spadesuit$}}

\renewcommand{\thefootnote}{\fnsymbol{footnote}}
\setcounter{footnote}{0}

\begin{titlepage}

\def\thefootnote{\fnsymbol{footnote}}

\begin{center}

\hfill March, 2022\\

\vskip .75in

{\Large \bf
  Observing Axion Emission from Supernova 
  \\[1mm]
  with Collider Detectors
  \\

}

\vskip .5in

{\large
  Shoji Asai$^{(a,b)}$,
  Yoshiki Kanazawa$^{(a)}$, Takeo Moroi$^{(a)}$\\[1mm]
  and Thanaporn Sichanugrist$^{(a)}$
  \\
}

\vskip 0.5in

$^{(a)}$
{\em
Department of Physics, The University of Tokyo, Tokyo 113-0033, Japan
}

\vskip 0.1in

$^{(b)}$
{\em
International Center for Elementary Particle Physics, 
The University of Tokyo, 
\\
Tokyo 113-0033, Japan
}

\end{center}
\vskip .5in

\begin{abstract}

  We consider a possibility to observe the axion emission from a
  nearby supernova (SN) in the future, which can be known in advance
  by the pre-SN alert system, by collider detectors like the LHC
  detectors (i.e., the ATLAS and the CMS) and the ILC detectors (i.e.,
  the ILD and SiD).  The axion from the SN can be converted to the
  photon by the strong magnetic field in the detector and the photon
  can be detected by electromagnetic calorimeter.  We estimate the
  numbers of signal and background events due to a nearby SN and show
  that the number of signal may be sizable.  The axion emission from a
  nearby SN may be observed if, at the time of the SN, the beam is
  stopped and the detector operation is switched to the one for the SN
  axion search.

\end{abstract}

\end{titlepage}

\renewcommand{\thepage}{\arabic{page}}
\setcounter{page}{1}
\renewcommand{\thefootnote}{\#\arabic{footnote}}
\setcounter{footnote}{0}
\renewcommand{\theequation}{\thesection.\arabic{equation}}

\section{Introduction}
\label{sec:intro}
\setcounter{equation}{0}

Axions, including the QCD axion in association with the spontaneous
breaking of the Peccei-Quinn symmetry \cite{Peccei:1977hh,
  Peccei:1977ur, Weinberg:1977ma, Wilczek:1977pj} as well as so-called
axion like particles (ALPs) arising from string theory
\cite{Svrcek:2006yi, Arvanitaki:2009fg, Cicoli:2012sz}, have been
attracted many attentions.  They are regarded as well-motivated
candidates of physics beyond the standard model and are important
targets of present and future experiments.  Axions are very weakly
interacting and the collider experiments (including beamdump and
fixed-target ones) can cover a very limited part of the parameter
space of the models.

In acquiring information about axions, astrophysical and cosmological
arguments are very powerful (for review, see, for example,
Ref.\ \cite{ParticleDataGroup:2020ssz} and references therein).
Particularly, paying attention to the coolings, various constraints on
axion parameters have been derived by the studies of the axion
emissions from stellar objects, like 
SN1987A \cite{Turner:1987by, Raffelt:2006cw, Chang:2018rso, Carenza:2019pxu}, and neutron stars \cite{Hamaguchi:2018oqw, Leinson:2021ety, Buschmann:2021juv}.
The cooling of other stellar objects, like white dwarfs, horizontal-branch stars, and red giant branch stars, have also provided astrophysical hints on the axion model \cite{Raffelt:2011ft, Ayala:2014pea, Giannotti:2015kwo, Giannotti:2017hny, Saikawa:2019lng, DiLuzio:2021ysg}.
Many of
these have accessed parameter regions which are out of the reaches of
collider experiments.

Because a significant amount of the axion may be emitted from stellar
objects, it is desirable to directly confirm the axion emission.  Such
possibilities have been considered in earlier literatures.
Ref.\ \cite{Engel:1990zd} pointed out that the axion from supernovae
(SNe) can excite the oxygen as $^{16}{\rm O}+a\rightarrow {^{16}{\rm
    O}}^*$ (with $a$ being the axion) in water C\v{e}renkov detectors
and that the photon emitted by the deexitation of ${^{16}{\rm O}}^*$
can be observed; negative observation in the Kamiokande detector at
the time of SN1987A ruled out a range of axion-nucleon-nucleon
coupling.  Then, in Ref.\ \cite{Moroi:1998qs}, it was claimed that, if
a new SN occurs in the future, a class of axion models predict
significant number of such signal events in SuperKamiokande detector
to confirm the model.  Another idea of the axion detection, so called
axion helioscope, is to convert solar axion to photon by using strong
magnetic field.  Currently, the most stringent bound on the
axion-photon-photon coupling is obtained by one of the axion
helioscope experiments, CERN Axion Solar Telescope (CAST)
\cite{CAST:2017uph}.  In addition, recently, Ref.\ \cite{Ge:2020zww}
argued that, if a detector sensitive to $\sim 100\ {\rm MeV}$ photon
is installed on a helioscope, detection of the axion from a nearby SN
in the future may be possible with the help of the pre-SN neutrino
alert \cite{Mukhopadhyay:2020ubs, Kato:2020hlc}.

In this article, we propose a new idea to detect axion emitted from a
SN in the future.  Our idea is to use particle detectors of collider
experiments for the detection of axion emitted from a nearby SN.  In
the central tracker region of collider detector, strong magnetic field
is applied.  Axion going through such a magnetic field can be
converted to the photon via the axion-photon-photon coupling.  The SN
axion has energy of $\sim O(10-100)\ {\rm MeV}$, so does the photon of
our interest.  Such photons can be detected particularly by the
electromagnetic calorimeter (ECAL) surrounding the tracker region. We
will show that, if a new SN occurs within $\sim O(100)\ {\rm pc}$, a
sizable number of axions are converted to the photons in collider
detectors.  There are SN progenitor candidates nearby; candidates
closer than $\sim 400\ {\rm pc}$ are listed in Table
\ref{table:candidates}~\cite{Mukhopadhyay:2020ubs}.  The occurrence of
a nearby SN can be alerted in advance.

The signal of the SN axion may be observed by collider detectors with
adopting special procedures at the time of the alert (which include
the turn off of the beam and the change of the detector operation to
the one dedicated for the SN axion search).  We propose each detector
collaboration to prepare in advance a detailed procedure for a nearby
SN.  Our proposal is low cost.  No new hardware component is
necessary, provided that the dedicated detector operation can be
realized at the software level.  The collider experiment can be
performed normally when there is no alert.

\begin{table}[t]
  \begin{center}
    \begin{tabular}{lll}
      \hline\hline
      HIP & Common Name & Distance (pc)
      \\
      \hline
      65474 & Spica / $\alpha$\,Virginis & $77 (4)$~\cite{vanLeeuwen:2007tv}
      \\
      81377 & $\zeta$\,Ophiuchi & $112 (2)$~\cite{vanLeeuwen:2007tv}
      \\
      71860 & $\alpha$\,Lupi & $143 (3)$~\cite{vanLeeuwen:2007tv}
      \\
      80763 & Antares / $\alpha$\,Scorpii & $169 (30)$~\cite{vanLeeuwen:2007tv}
      \\
      107315 & Enif / $\epsilon$\,Pegasi & $211 (6)$~\cite{vanLeeuwen:2007tv}
      \\
      27989 & Betelgeuse / $\alpha$\,Orionis & $222_{-34}^{+48}$~\cite{Harper_2017}
      \\
      109492 & $\zeta$ Cephei & $256 (6)$~\cite{Brown2018GaiaDR}
      \\
      24436 & Rigel / $\beta$ Orionis & $264(24)$~\cite{vanLeeuwen:2007tv}
      \\
      31978 & S Monocetotis A(B) & $282(40)$~\cite{vanLeeuwen:2007tv}
      \\
      25945 & CE Tauri / 119 Tauri & $326(70)$~\cite{Brown2018GaiaDR}
      \\
      \hline\hline
    \end{tabular}
    \caption{Candidates of SN progenitor, which are both red- and blue-supergiants with a mass larger than $\sim 10~M_{\odot}$~\cite{Mukhopadhyay:2020ubs}. The Hipparcos catalog number is shown in the first column. }
    \label{table:candidates}
  \end{center}
\end{table}

This article is organized as follows.  In Section \ref{sec:SNaxion},
we summarize the axion emission from the SN.  In Section
\ref{sec:conversion}, we discuss the axion conversion to photon in
collider detectors.  The procedure to detect the signal of the SN
axion at collider detectors is discussed in Section
\ref{sec:detection}.  Section \ref{sec:conclusions} is dedicated for
conclusions and discussion.

\section{Axion and SN}
\label{sec:SNaxion}
\setcounter{equation}{0}

In this Section, we overview basic properties of the axion and its
emission from the SN.  The relevant part of the Lagrangian for our
discussion is given by
\begin{align}
  {\cal L} = {\cal L}_{\rm SM}
  + \frac{1}{2} \partial_\mu a \partial^\mu a - \frac{1}{2} m_a^2 a^2
  + \frac{1}{4} g_{a\gamma\gamma} a F_{\mu\nu} \tilde{F}^{\mu\nu}
  + \sum_{N=p,n} \frac{g_{aNN}}{2m_N} \bar{N}
  \gamma^\mu \gamma_5 N \partial_\mu a,
  \label{Lagrangian}
\end{align}
where ${\cal L}_{\rm SM}$ is the standard-model Lagrangian,
$F_{\mu\nu}\equiv\partial_\mu A_\mu-\partial_\nu A_\mu$ (with $A_\mu$
being the vector potential of the photon) is the field strength tensor
of the electromagnetic field, and $m_N$ is the nucleon mass.  The
coupling constants $g_{a\gamma\gamma}$, $g_{app}$, and $g_{ann}$ are
model-dependent.  Because we consider a wide class of axion models, we
treat $g_{a\gamma\gamma}$, $g_{app}$, and $g_{ann}$ as free
parameters.  For the case of the QCD axion, for example, these
parameters are inversely proportional to the axion decay constant
$f_a$ as
\begin{align}
  g_{a\gamma\gamma} = \frac{1}{f_a}C_{a\gamma\gamma},~~~
  g_{app} = \frac{m_N}{f_a}C_{app},~~~
  g_{ann} = \frac{m_N}{f_a}C_{ann}.
  \label{gaxx}
\end{align}
The coefficients, as well as the mass of the QCD axion, are precisely
calculated in Ref.\ \cite{GrillidiCortona:2015jxo}.  The coefficient
of the axion-photon-photon coupling is given by
\begin{align}
  C_{a\gamma\gamma} \simeq
  \frac{\alpha}{2\pi} \left( \frac{E}{N} - 1.92 \right),
\end{align}
where $\alpha$ is the fine structure constant and $E/N$ is the ratio
of the the electromagnetic and color anomalies of the axial current
associated with the axion.  The coefficients of the
axion-nucleon-nucleon coupling for the KSVZ axion
\cite{Kim:1979if,Shifman:1979if} is given by
\begin{align}
  C_{app}^{\rm (KSVZ)} \simeq 0.47,~~~
  C_{ann}^{\rm (KSVZ)} \simeq 0.02, 
\end{align}
while for the case of DFSZ axion \cite{Dine:1981rt,
  Zhitnitsky:1980tq}, the coefficients are
\begin{align}
  C_{app}^{\rm (DFSZ)} \simeq -0.617 + 0.435 \sin^2 \beta,~~~
  C_{ann}^{\rm (DFSZ)} \simeq  0.254 - 0.414 \sin^2 \beta,
  \label{gann_DFSZ}
\end{align}
with $\tan\beta$ being the ratio of the vacuum expectation values of
two Higgs doublets.  The mass of the QCD axion is related to $f_a$ as
\begin{align}
  m_a^{\rm (QCD)} \simeq 
  5.70\ {\rm meV} \times
  \left( \frac{f_a}{10^{9}\ {\rm GeV}} \right)^{-1}.
\end{align}

In the core of the SN, the axion can be produced by the scattering
processes.  We mostly consider the case that the axion emission is so
ineffective that it does not significantly affect the cooling of the
SN.  Then, the cooling process is dominated by the neutrino emission.
In the following, the axion emission rate for such a case will be
considered adopting the results of Refs.\ \cite{Carenza:2019pxu,
  Carenza:2020cis}.  We will comment on the case that the axion
emission rate becomes comparable to the neutrino emission rate later.

\begin{figure}[t]
  \centering
  \includegraphics[width=0.7\textwidth]{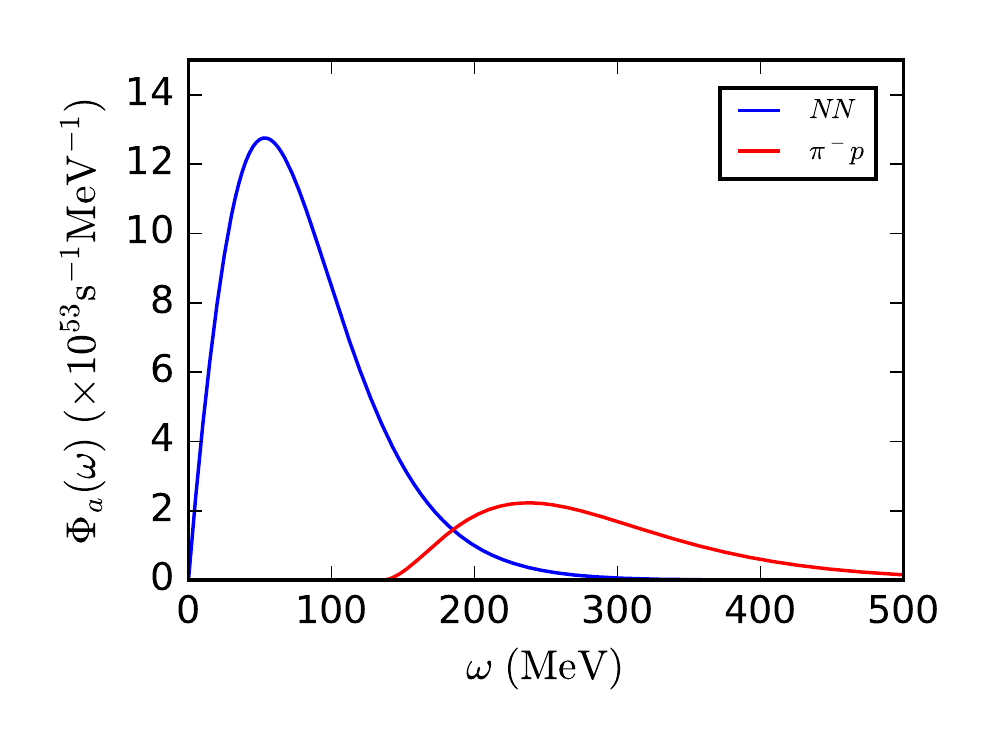}
  \caption{The spectra of the axion emitted from a SN, taking $\tilde{g}_{aNN}=5 \times10^{-10}$, $\kappa_{\rm SN}=3$, and
    $L_a^{(\pi^- p)}=L_a^{(NN)}$.
    The blue line is for the $NN$ process, which is calculated in our SN 
    model. The red line is for the $\pi^{-1}p$ process, which is obtained 
    from the fitting formula shown in Eq.\ \eqref{Phia(pip)}.
    }
  \label{fig:Phi}
\end{figure}

One of the important axion emission processes is the $NN$
bremsstrahlung process:
\begin{align}
  N_1 + N_2 \rightarrow N_1 + N_2 + a,
\end{align}
where $N_i$ ($i=1,2$) are nucleons (i.e., proton $p$ or neutron $n$).
We adopt the analysis of Ref.\ \cite{Carenza:2019pxu} to estimate the
axion emission rate due to the $NN$ bremsstrahlung process.  The axion
luminosity mildly depends on time when $t_{\rm pb}\lesssim 10\ {\rm
  sec}$ (with $t_{\rm pb}$ being the post-bounce time).  At $t_{\rm
  pb}=1\ {\rm sec}$, the axion luminosity due to the $NN$
bremsstrahlung process
is given by 
$2.42 \times 10^{70}\ {\rm erg/sec} \times \tilde{g}_{aNN}^2$, 
where $\tilde{g}_{aNN}$ is the effective axion-nucleon-nucleon coupling
constant defined as
\begin{align}
  \tilde{g}_{aNN}^2 \equiv
  g_{ann}^2 + 0.61 g_{app}^2 + 0.53 g_{ann}g_{app}.
\end{align}
Then, the luminosity due to the bremsstrahlung process gradually
increases with time until $t_{\rm pb}\sim 5\ {\rm sec}$ and is a few
times larger than that at $t_{\rm pb}=1\ {\rm sec}$ when $t_{\rm
  pb}\lesssim 10\ {\rm sec}$.  Based on these observations, we
parameterize the time-averaged axion luminosity due to the
bremsstrahlung process as
\begin{align}
  L_a^{(NN)} =
  \int d\omega \omega \Phi_a^{(NN)} (\omega) =  
  2.42 \times 10^{70}\ {\rm erg/sec} \times \tilde{g}_{aNN}^2 
  \kappa_{\rm SN},
  \label{La(NN)}
\end{align}
where $\Phi_a^{(NN)}$ is the time-averaged axion flux, and
$\kappa_{\rm SN}$, which is expected to be a few, is a numerical
factor to take into account the effect of the increase of the
luminosity after $t_{\rm pb}=1\ {\rm sec}$.  For the calculation of
$\Phi_a^{(NN)}$, we follow the analysis of Ref.\ \cite{Carenza:2019pxu}
to determine the shape of the spectrum while the normalization is
fixed by Eq.\ \eqref{La(NN)}. We adopt the simple uniform SN model
with the proton fraction $Y_p=0.3$, the density $\rho=1.3
\times10^{14}~\rm{g/cm^3}$, and the temperature $T_{\rm
  SN}=35~\rm{MeV}$ for the KSVZ axion model.  The calculation is
performed beyond one-pion-exchange approximation including medium
corrections. In Fig.\ \ref{fig:Phi}, we plot the axion spectrum
$\Phi_a^{(NN)}(\omega)$ as a function of $\omega$, taking
$\tilde{g}_{aNN}=5 \times10^{-10}$ and $\kappa_{\rm SN}=3$ (blue
line). The peak of the axion spectrum due to the $NN$ scattering is at
$\omega\sim 70 \ \rm{MeV}$.

As well as the $NN$ bremsstrahlung process, the $\pi^- p$ scattering
process also produces the axion efficiently:
\begin{align}
  \pi^- + p\rightarrow n + a.
\end{align}
In many literatures, it has been expected that the axion production in
the SN is dominated by the $NN$ bremsstrahlung process.  However,
recently, Refs.\ \cite{Carenza:2020cis,Fischer:2021jfm} pointed out
that the $\pi^- p$ scattering process is also important.  The axion
emissivity due to the $\pi^- p$ scattering process is comparable to
that due to the $NN$ bremsstrahlung process and the axion spectrum
from the $\pi^- p$ process is harder than that from the $NN$
bremsstrahlung process.  For the $\pi^- p$ process, we assume that the
axion spectrum is proportional to the one given in
Ref.\ \cite{Carenza:2020cis}, which we found can be well approximated
as
\begin{align}
  \Phi_a^{(\pi^- p)} (\omega) \propto
  (\omega-135\ {\rm MeV})^{2.07}\, e^{-0.0202\, (\omega/1\ {\rm MeV})}.
  \label{Phia(pip)}
\end{align}
Because the axion luminosities due to the $NN$ and $\pi^- p$ processes
are expected to be of the same order \cite{Carenza:2020cis}, we simply
assume that the time-averaged luminosity due to the $\pi^- p$ process,
denoted as $L_a^{(\pi^- p)}$, is equal to that due to the $NN$
bremsstrahlung process to determine the normalization of
$\Phi_a^{(\pi^- p)}$.  We use Eq.\ \eqref{Phia(pip)} to determine the
(time-averaged) spectrum of the axion from the $\pi^- p$ process.  In
Fig.\ \ref{fig:Phi}, we also show $\Phi_a^{(\pi^- p)}$, taking
$L_a^{(\pi^- p)}=L_a^{(NN)}$.  We can see that the spectrum is peaked
at $\omega\sim 240\ {\rm MeV}$.  Notice that, because the axion from
the $\pi^- p$ process has harder spectrum, the number of axion from
the $\pi^- p$ process is smaller than that from the $NN$
bremsstrahlung process if $L_a^{(\pi^- p)}\sim L_a^{(NN)}$.

The total number of axion emitted from the SN is calculated as
\begin{align}
  N_a=\dot{N}_a \Delta t_{\rm SN},
\end{align}
where $\Delta t_{\rm SN}$ is the post-bounce time when the axion
emission becomes ineffective, and $\dot{N}_a$ is the axion emission
rate:
\begin{align}
  \dot{N}_a \equiv \int d\omega \Phi_a (\omega),
\end{align}
with $\Phi_a\equiv \Phi_a^{(NN)} + \Phi_a^{(\pi^-p)}$.

Before closing this section, we summarize constraints on the axion
couplings which will be used in the following discussion.  In order
not to affect the SN cooling too much, the effective
axion-nucleon-nucleon coupling is bounded from above.  In
Ref.\ \cite{Carenza:2019pxu}, the bound is given by $\tilde{g}_{aNN}
\lesssim 9.1 \times 10^{-10}$, assuming that the axion production is
dominated by the $NN$ bremsstrahlung process.  Here, because we also
consider the $\pi^- p$ process with taking $L_a^{(\pi^- p)}\sim
L_a^{(NN)}$, we adopt the following bound:
\begin{align}
  \tilde{g}_{aNN} \lesssim 6.4 \times 10^{-10}.
  \label{bound_gann}
\end{align}
The axion-photon-photon coupling $g_{a\gamma\gamma}$ is also bounded
from above.  The most stringent constraint on the axion-photon-photon
coupling is given by the non-observation of the axion emission from
the sun.  The CAST experiment provides the upper bound on
$g_{a\gamma\gamma}$ as \cite{CAST:2017uph}
\begin{align}
  g_{a\gamma\gamma} \lesssim 0.66 \times 10^{-10}\ {\rm GeV}^{-1}.
  \label{bound_gagg}
\end{align}

\section{Axion Conversion to Photon}
\label{sec:conversion}
\setcounter{equation}{0}

Next, let us briefly review the conversion of the axion to photon in
the presence of magnetic field \cite{Sikivie:1983ip, Sikivie:1985yu,
  Raffelt:1987im}.  Such a conversion process plays an essential role
in observing the SN axion with collider detectors.

We are interested in the propagation of plane-wave axion and photon
under the influence of static and uniform magnetic field.  In such a
situation, axion mixes with the photon whose polarization vector is in
the plane containing the propagation direction and the direction of
the magnetic field.  (The amplitude of the photon with such a
polarization is denoted as $A_{||}$).  Considering the mode with the
oscillation frequency $\omega$, Lagrangian given in
Eq.\ \eqref{Lagrangian} results in the following evolution equation of
the axion $a$ and $A_{||}$ \cite{Raffelt:1987im}:\footnote
{With the external magnetic field, the refractive indices of the
  photon are affected.  In addition, if we consider the
  propagation in a matter, photon may acquire effective mass
  (so-called plasma frequency).  We consider the case in which these
  effects are negligible.}
\begin{align}
  \left( 
  \begin{array}{cc}
    \omega^2 + \partial_\lpath^2 & g_{a\gamma\gamma} B_T \omega \\
    g_{a\gamma\gamma} B_T \omega & \omega^2 + \partial_\lpath^2 -m_a^2
  \end{array}
  \right)
  \left( 
  \begin{array}{c}
    A_{||} \\ a
  \end{array} 
  \right) = 0,
\end{align}
where $\lpath$ denotes the propagation distance, and $B_T \equiv
B\sin\theta$ with $\theta$ being the angle between the propagation
direction and the direction of the magnetic field.  (In collider
detectors, the magnetic field is parallel to the beam axis and hence
$\theta$ is the angle between the beam axis and the direction of the
SN from the earth.)  Using the fact that the dispersion relations of
$a$ and $A_{||}$ are well approximated by $k\simeq \omega$ (with $k$
being the wave number), the above equation can be expressed as
\begin{align}
  i\partial_\lpath 
  \left(
  \begin{array}{c}
    A_{||} \\ a
  \end{array} 
  \right) =
  \left( 
  \begin{array}{cc}
    \omega & g_{a\gamma\gamma} B_T / 2 \\
    g_{a\gamma\gamma} B_T /2 & \omega - q
  \end{array}
  \right)
  \left( 
  \begin{array}{c}
    A_{||} \\ a
  \end{array}
  \right),
\end{align}
where 
\begin{align}
  q \equiv \frac{m_a^2}{2\omega}.
\end{align}
Numerically, $q^{-1}\simeq 4\ {\rm m}\times (\omega/10\ {\rm MeV})
(m_a/1\ {\rm eV})^{-2}$.

Solving the above differential equation, the conversion probability from
the axion to the photon is given by
\begin{align}
  P (\lpath) = \frac{1}{4} (g_{a\gamma\gamma} B_T \lpath)^2
  \left( \frac{\sin (q\lpath/2)}{q\lpath/2} \right)^2.
  \label{prob}
\end{align}
Notice that the conversion probability grows as $\lpath^2$ as far as
$\lpath\lesssim q^{-1}$.

Now, we consider the conversion of the SN axion to the photon in
collider detectors.  Hereafter, we consider a detector with the
conventional design such that a strong magnetic field is applied in
the central region where detectors for the tracking are installed.  We
estimate the number of the photon converted from the SN axion assuming
that the density of the material in the central region is so small
that the scattering probability of $O(10-100)\ {\rm MeV}$ photon is
negligible.  Thus, in the central region of the detector, the photon
is assumed to propagate (almost) freely; validity of this assumption
will be considered later.  The photon converted from the SN axion is
expected to be observed by the ECAL surrounding the central region.

In calculating the number of photon converted from the SN axion, for
simplicity, we approximate that the shape of the central region as a
cylinder with the radius of $R$ and the length of $\Ldet$ and that the
magnetic field in the central region is uniform.  Here, we consider
the ATLAS and the CMS detectors of the LHC experiment and the ILD and
the SiD detectors of the ILC experiment.\footnote
{We have also estimated the number of events for the Belle II detector
  of the SuperKEKB experiment.  For the Belle II detector, $N_\gamma$
  is at most $O(1)$ in the parameter region which is currently viable,
  and hence the SN axion is hardly observed by the Belle II detector.}
The parameters characterizing these detectors are summarized in Table
\ref{table:detectors}.\footnote
{Currently, the ATLAS detector has Transition Radiation Tracker (TRT)
  at $50\lesssim r\lesssim 110\ {\rm cm}$ (with $r$ being the distance
  from the beam pipe).  The photon may not be treated as a free
  particle in the TRT region where the photon is converted to $e^+e^-$
  pair with a significant probability.  During the Long Shutdown 3 after
  Run 3, TRT (and other inner detectors) will be replaced by
  all-silicon inner tracker, for which the conversion probability is
  expected to be relatively low.  The $R$ parameter of the ATLAS
  detector given in Table \ref{table:detectors} is the value after
  the Long Shutdown 3.}

\begin{table}[t]
  \begin{center}
    \begin{tabular}{l|cccccc}
      \hline\hline
      & $R$ & $\Ldet$ & $B$
      & $\bar{\lpath} (\theta=\frac{\pi}{6})$
      & $\bar{\lpath} (\theta=\frac{\pi}{3})$
      & $\bar{\lpath} (\theta=\frac{\pi}{2})$
      \\
      \hline
      ATLAS (LHC) \cite{ATLAS:1999uwa}
      & $1.1\ {\rm m}$ & $6.7\ {\rm m}$ & $2.0\ {\rm T}$
      & $1.2\ {\rm m}$ & $1.4\ {\rm m}$ & $1.8\ {\rm m}$
      \\
      CMS (LHC) \cite{CMS:2006myw}
      & $1.3\ {\rm m}$ & $5.8\ {\rm m}$ & $3.8\ {\rm T}$
      & $1.3\ {\rm m}$ & $1.9\ {\rm m}$ & $2.1\ {\rm m}$
      \\
      ILD (ILC) \cite{Behnke:2013lya}
      & $1.8\ {\rm m}$ & $4.9\ {\rm m}$ & $3.5\ {\rm T}$
      & $1.6\ {\rm m}$ & $2.6\ {\rm m}$ & $2.9\ {\rm m}$
      \\
      SiD (ILC) \cite{Behnke:2013lya}
      & $1.2\ {\rm m}$ & $3.3\ {\rm m}$ & $5.0\ {\rm T}$
      & $1.1\ {\rm m}$ & $1.7\ {\rm m}$ & $2.0\ {\rm m}$
      \\
      \hline\hline
    \end{tabular}
    \caption{Detector parameters adopted in our analysis.  The
      magnetic field of the ATLAS detector in the tracking volume
      deviates significantly from uniformity \cite{ATLAS:1999uwa}; we
      take $B=2.0\ {\rm T}$ as an averaged magnetic field.  We also
      show the effective path length for $\theta=\frac{\pi}{6}$,
      $\frac{\pi}{3}$, and $\frac{\pi}{2}$, for the case of
      $q\bar{\lpath}\ll 1$.}
    \label{table:detectors}
  \end{center}
\end{table}

For a given SN, the spectrum of the photon converted from the SN axion
is given in the following form:
\begin{align}
  F_{\gamma}  (\omega) = 
  \frac{\Delta t_{\rm SN}}{4\pi d_{\rm SN}^2}
  \Phi_a (\omega)
  \int dA P (\lpath_A).
\end{align}
The integration is over the cross-section area.  The propagation
length of the path inside the cylinder (i.e., the central region of
the detector) is dependent on the cross-section area and is denoted as
$\lpath_A$; $P (\lpath_A)$ is the conversion probability
for such a path.\footnote
{Cross section area for a given plane-wave axion can be
  parameterized as
  \begin{align*}
    -R \leq x' \leq R, ~~~ 
    -z'_{\rm max} \leq z' \leq z'_{\rm max},
  \end{align*}
  where $z'_{\rm max} \equiv \frac{1}{2} \Ldet \sin\theta + \sqrt{R^2
    - x'^2} \cos\theta$.  The integration can be understood as
  \begin{align*}
    \int dA P (\lpath_A) \equiv 
    \int_{-R}^R dx' \int_{-z'_{\rm max}}^{z'_{\rm max}} dz'
    P (\lpath(x',z')).
  \end{align*}
  Here, $\lpath_A=\lpath(x', z')$ is the maximal propagation length
  inside the cylinder for the path going through $(x', z')$, and is given by
  \begin{align*}
    \lpath(x', z') = 
    \left\{
    \begin{array}{ll}
      2\sqrt{R^2 - x'^2} / \sin\theta &
      : \mbox{if}~ z'_{\rm th} > 0 ~\mbox{and}~ |z'| < |z'_{\rm th}| 
      \\[1mm]
      \Ldet / \cos\theta &
      : \mbox{if}~ z'_{\rm th} < 0 ~\mbox{and}~ |z'| < |z'_{\rm th}| 
      \\[1mm]
      (z'_{\rm max} - |z'|) / \sin\theta \cos\theta & 
      : \mbox{if}~ |z'| > |z'_{\rm th}|
    \end{array}
    \right. ,
  \end{align*}
where $z'_{\rm th} \equiv \frac{1}{2} \Ldet \sin\theta - \sqrt{R^2 -
  x'^2} \cos\theta$.}

In order to discuss the axion-photon conversion in detectors, we
define the ``effective path length'' as follows:
\begin{align}
  \bar{\lpath} \equiv
  \left[
    \frac{1}{2 R \Ldet}
    \int dA\, \lpath_A^2
    \left( \frac{\sin (q\lpath_A/2)}{q\lpath_A/2} \right)^2
    \right]^{1/2}
  \sin\theta.
\end{align}
Notice that $2R\Ldet$ is the total cross section area for
$\theta=\frac{\pi}{2}$.  In Table \ref{table:detectors},
$\bar{\lpath}$ is also shown for several values of $\theta$ taking
$q\rightarrow 0$.  Using $\bar{\lpath}$, the number of photon
converted from the SN axion can be expressed as
\begin{align}
  N_{\gamma} =\, 
  \frac{\dot{N}_a \Delta t_{\rm SN}}{8\pi d_{\rm SN}^2}
  R \Ldet (g_{a\gamma\gamma} B \bar{\lpath})^2.
  \label{N_gamma}
\end{align}
Numerically, we find 
\begin{align}
  N_{\gamma} \simeq &\, 
  25 \times
  \left( \frac{2 R \Ldet}{15\ {\rm m^2}} \right)
  \left( \frac{\bar{\lpath}}{2\ {\rm m}} \right)^2
  \left( \frac{B}{4\ {\rm T}} \right)^2
  \nonumber \\ &\, \times
  \left( \frac{\kappa_{\rm SN}}{3} \right)  
  \left( \frac{d_{\rm SN}}{100\ {\rm pc}} \right)^{-2}
  \left( \frac{\Delta t_{\rm SN}}{10\ {\rm sec}} \right)
  \left( \frac{g_{a\gamma\gamma}}{7\times 10^{-11}\ {\rm GeV}^{-1}} \right)^{2}
  \left( \frac{\tilde{g}_{aNN}}{6 \times 10^{-10}} \right)^{2}.
  \label{N_gamma(numerical)}
\end{align}

From Eq.\ \eqref{N_gamma(numerical)}, we can see that the number of
the photon converted from the SN axion may be significant.  Taking
$g_{ann}=6.4\times 10^{-10}$, $g_{a\gamma\gamma}=6.6\times
10^{-11}\ {\rm GeV}^{-1}$, which correspond to the maximal possible
values of the effective axion-nucleon-nucleon coupling and
axion-photon-photon coupling, respectively (see
Eqs.\ \eqref{bound_gann} and \eqref{bound_gagg}), $d_{\rm SN}=77\ {\rm
  pc}$ (corresponding to the distance to Spica), $\kappa_{\rm SN}=3$,
$\Delta t_{\rm SN}=10\ {\rm sec}$, and
$\theta=\frac{\pi}{6}-\frac{\pi}{2}$, $N_\gamma$ can be as large as
$4-8$, $16-43$, $25-81$, and $11-35$ for the ATLAS, CMS, ILD, and SiD
setups, respectively.  (For the case of the CMS, however, the
axion-photon conversion may be affected by the materials in the
central region, resulting in the suppression of the number of the
signal; see the discussion below.)

We comment here that the signal photon converted from the SN axion
enters only into (almost) a half of the ECAL.  In our procedure, the
ECAL can be geometrically divided into ``upper'' and ``lower'' halfs.
The upper half is defined as the ECAL where the SN axion passes before
entering into the central region, while the lower half is the rest of
the ECAL.  Then, the signal photon converted from the SN axion can be
detected only in the lower half.  This can be used to understand the
background properties as we will discuss later.

\section{Detection of the Signal of SN Axion}
\label{sec:detection}
\setcounter{equation}{0}

Now, we study the possibility of detecting the signal of the SN axion
at collider detectors.  It is highly non-trivial to identify the SN
axion signal during the normal operation of the collier experiments.
The signal photon may not be distinguished from photon (and other
particles) produced by the beam collision.  Thus a special procedure
is necessary for the detection of the SN axion at the time of the SN,
including the turn off of the beam.

For the detection of the SN axion signal, we use the fact that nearby
SNe are expected to be alerted well in advance by the pre-SN neutrino
alert.  A large number of neutrinos are emitted at the time of the
core collapse SN.  The neutrino flux is regularly monitored by
detectors which are sensitive to the SN neutrino, like KamLAND
\cite{KamLAND:2015dbn}, SNO+ \cite{SNO:2015wyx}, and SuperKamiokande
\cite{Super-Kamiokande:2019xnm}.  In the future, sensitivity to the
nearby SNe can be strengthened by observatories like JUNO
\cite{JUNO:2015sjr}, Hyper-Kamiokande \cite{Hyper-Kamiokande:2018ofw},
and DUNE \cite{DUNE:2016hlj, DUNE:2015lol, DUNE:2016evb,
  DUNE:2016rla}.  These observatories form a global network to provide
a prompt warning of a nearby SN when an anomalous increase of the
neutrino flux, which indicates a nearby SN, is found.  This global
network is called the Supernova Early Warning System
(SNEWS)~\cite{Antonioli:2004zb, Scholberg:2008fa, Vigorito:2011zza}.
The calculation of the neutrino flux 
\cite{Kato:2020hlc, Odrzywolek:2003vn, Odrzywolek:2004em,
  2009AcPPB..40.3063K, Odrzywolek:2009wa, Kato:2015faa,
  Patton:2015sqt, Yoshida:2016imf, Patton:2017neq, Kato:2017ehj,
  Guo:2019orq} 
shows that, if a SN progenitor is located within $\sim1~\rm{kpc}$, the
neutrino emission is large enough to be detected even before the
SN core collapse.  Such an increase of the pre-SN neutrino
makes it possible to forecast the SN which occur within $\sim
O(100)\ {\rm pc}$; for $d_{\rm SN}\lesssim 200\ {\rm pc}$, an alert is
possible $O(1-10)\ {\rm hours}$ before the core collapse
\cite{Mukhopadhyay:2020ubs, Kato:2020hlc}.  With the help of the
pre-SN alert, we may know when the collider operation should be
switched to the special one for the SN axion.

In the following, we first consider backgrounds.  Then, we discuss the
detectability of the signal at the ILC and LHC detectors.

\subsection{Backgrounds}

The most serious source of the background is expected to be the
neutrino emitted by the SN of our interest.  The neutrino emitted by
the nearby SN should arrive simultaneously with the SN axion and it
interacts with the material in the detector.  Most importantly, the
electron neutrino $\nu_e$ emitted from the SN interacts with heavy
nuclei in the calorimeter via the charged current interaction and
becomes electron.  Such electron produced in the ECAL may mimic the
signal photon.\footnote
{The charged current process induced by the SN neutrino may also be a
  serious background for the SN axion detection with the
  supernova-scope proposed in Ref.\ \cite{Ge:2020zww}.}

For the estimation of the background rate, we use the fitting formula
for the electron neutrino flux given in Ref.\ \cite{Tamborra:2012ac}:
\begin{align}
  f_{\nu_e} (E_\nu) =
  \frac{L_{\nu_e}}{4\pi d_{\rm SN}^2 \bar{E}_{\nu_e}^2}
  \frac{(\alpha_e+1)^{\alpha_e+1}}{\Gamma(\alpha_e+1)}
  \left(\frac{E_\nu}{\bar{E}_{\nu_e}}\right)^{\alpha_e} 
  e^{-(\alpha_e+1)E_\nu/\bar{E}_{\nu_e}}.
\end{align}
Here, $L_{\nu_e}$ is the total luminosity, $\alpha_e$ is a positive
constant of $O(1)$, and $\bar{E}_{\nu_e}$ is the averaged energy which
is $\sim 10-15\ {\rm MeV}$.  These parameters depend on time and we 
adopt the parameters for $t_{\rm pb}=1.991\ {\rm sec}$, i.e.,
$L_{\nu_e}=4.8\times 10^{51}\ {\rm erg/sec}$, $\alpha_e=2.92$, and
$\bar{E}_{\nu_e}=10.01\ {\rm MeV}$, to evaluate the background
rate.\footnote
{The charged current scattering cross section of $\bar{\nu}_e$ is
  order of magnitude smaller than that of $\nu_e$ (at least for Pb and
  Fe) \cite{Kolbe:2000np}.  Thus, we neglect the effect of
  $\bar{\nu}_e$.}

As we will see below, the scattering rate due to the SN neutrino is
much higher than the signal rate (which is at most $O(10)\ {\rm Hz}$).
Thus, we should somehow remove the background due to the SN neutrino.
For this purpose, we propose to use the difference of energy
distributions of the SN axion and the SN neutrino.  The signal photon
and the background electron inherit the energy of the parent axion and
neutrino, respectively.  (In our analysis, for simplicity, the energy
of the electron produced by the charged current process is
approximated to be equal to that of the parent SN neutrino.)  The
energy of the electron produced by the SN neutrino is typically $\sim
10\ {\rm MeV}$, while the typical energy of the signal photon is
higher.  With imposing a cut on the energy deposit in the ECAL, we may
remove the background.

The cosmic-ray muons may also leave some activity in the calorimeter.
We expect that the muon detector, which is located at the most outer
region of the detector, can be used as the veto counter to remove the
cosmic-ray background.  Thus, we neglect such a background.

\subsection{Detection at the ILC detectors}

First, we consider the ILC detectors which provide better environment
for the SN axion detection than the LHC ones.  In particular, we
consider the case with the ILD, which has longer effective path length
than the SiD, in detail.  In the case of the ILD, the total radiation
length up to the outside of the central tracker region is $\sim 0.1$
\cite{Behnke:2013lya}, and we may regard the photon as a freely
propagating particle in the central region.

The ILC detectors are planned to have tungsten-based ECALs.  For the
ILD setup, the total depth of the tungsten is $8.4\ {\rm cm}$ (i.e.,
$24$ radiation length) to the radial direction, corresponding to the
total weight of $\sim 120\ {\rm t}$ (equivalent to $\sim 4\times
10^{29}$ of tungsten atom).  The SN neutrino, in particular, $\nu_e$,
interacts with the tungsten in the ECAL via the charged current
interaction.  We could not find the cross section for such a
background process, $\nu_e({\rm W},e^-)X$, while the cross section of
$\nu_e$ with lead, $\nu_e({\rm Pb},e^-)X$, can be found in
Ref.\ \cite{Tamborra:2012ac,Kolbe:2000np} and is of order
$10^{-41}-10^{-39}\ {\rm cm}^2$ for the neutrino energy of our
interest.  Based on the observation that the neutrino scattering cross
section with nuclei tends to increase as the atomic number becomes
larger, we presume that the scattering cross section for $\nu_e({\rm
  W},e^-)X$ is at most comparable to (or smaller than) that for
$\nu_e({\rm ^{208}Pb},e^-)X$.  In estimating the background rate, we
take the cross section for $\nu_e({\rm W},e^-)X$ to be equal to that
for $\nu_e({\rm ^{208}Pb},e^-)X$.  We expect that such an
approximation gives a conservative estimation of the background rate.

\begin{figure}[t]
  \centering
  \includegraphics[width=0.7\textwidth]{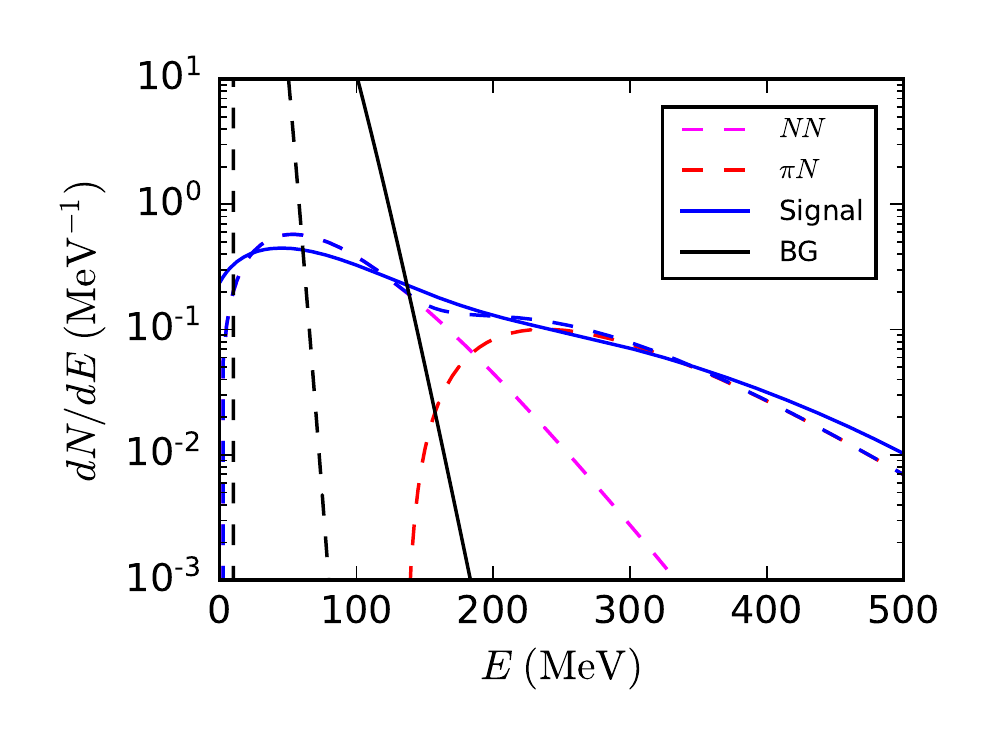}
  \caption{The spectrum of the signal and background events for the
    setup of the ILD, taking $\tilde{g}_{aNN}=6.4 \times 10^{-10}$,
    $g_{a\gamma\gamma}=6.6\times 10^{-11}\ {\rm GeV}^{-1}$, $d_{\rm
      SN}=77\ {\rm pc}$, $\Delta t_{\rm SN}=10\ {\rm sec}$,
    $\kappa_{\rm SN}=3$, $\theta=\frac{\pi}{2}$, and
    $q^{-1}\gg\bar{L}$. Signal and background spectrum are shown in
    blue and gray lines, respectively. Solid lines represent the
    spectrum with taking into account the detector effect, while
    dashed lines represent non-smeared spectra. }
  \label{fig:dnde_ILD}
\end{figure}

In Fig.\ \ref{fig:dnde_ILD}, we show in dashed lines the spectrum of
the photon converted from the SN axion, as well as that of the
electron due to the SN neutrino, for the ILD setup.\footnote
{In the estimation of the background induced by the SN neutrino, we count the total number of event in the whole lower half of the ECAL. The background due to the SN neutrino may be reduced by using the information about the event location.  In the baseline design, the ILD ECAL is segmented into 30 layers. The signal photon converted from the SN axion should go through the most inner layer while the background event due to the SN neutrino can occur at any place of the ECAL.  Requiring a large energy deposit in inner layers, we may omit a sizable amount of background.  If only the SN neutrino event which occur in the most inner layer contributes to the background, for example, the number of background is reduced by the factor of $\sim 30$.  In such a case, we can take a lower value of $E_{\rm cut}$ and $N_{\rm signal}$ can be increased by $\sim 30\ \%$.}
We show the total signal with blue (dashed) line and the backgrounds
with gray (dashed) line, while separately show the contribution from
$NN$ bremsstrahlung process and pion-induced process. Here, we use the
maximal possible values of $\tilde{g}_{aNN}=6.4 \times 10^{-10}$ and
$g_{a\gamma\gamma}=6.6\times 10^{-11}\ {\rm GeV}^{-1}$, and take
$d_{\rm SN}=77\ {\rm pc}$, $\Delta t_{\rm SN}=10\ {\rm sec}$,
$\kappa_{\rm SN}=3$ with $\theta=\pi/2$.  With such a choice of
parameters, the total number of the scattering event by the SN
neutrino is $\sim 10^5$ and hence the rate is $\sim 10\ {\rm kHz}$.

Even though the number of the electron produced by the SN neutrino
exponentially decreases when $E\gtrsim 10\ {\rm MeV}$ (with $E$ being
the energy deposit in the ECAL), the observed background spectrum
should have longer high energy tail because of the energy resolution
of the ECAL.  Here, we estimate the observed spectrum adopting the
following energy resolution of the ILD detector \cite{Behnke:2013lya}:
\begin{align}
  \frac{\delta E}{E} = \frac{15  \%}{\sqrt{E_{\rm GeV}}},
  \label{de(ILD)}
\end{align}
where $E_{\rm GeV}$ is the energy in units of GeV.
The smeared signal and background spectra are also shown in
Fig.\ \ref{fig:dnde_ILD}.  We can see that the number of background can be
significantly reduced if we require large enough $E$.  We introduce
$E_{\rm cut}$ such that
\begin{align}
  \int_{E_{\rm cut}}^\infty dE
  \left[ \frac{dN_{\rm BG}}{dE} \right]_{\rm smeared} = 1,
  \label{Ecut}
\end{align}
where $[dN_{\rm BG}/dE]_{\rm smeared}$ is the background spectrum
after taking into account the detector resolution; for the present choice of
parameters, $E_{\rm cut}\simeq 145\ {\rm MeV}$. 
Then, we regard the region $E>E_{\rm cut}$ as the signal region and define
the number of signal as
\begin{align}
  N_{\rm signal} \equiv 
  \int_{E_{\rm cut}}^\infty dE
  \left[ \frac{dN_{\rm signal}}{dE} \right]_{\rm smeared},
\end{align}
with $[dN_{\rm signal}/dE]_{\rm smeared}$ being the photon spectrum
with the effect of the detector resolution included.\footnote
{The difference of the time curves of the axion and neutrino emissions
  may be also used to reduce the neutrino background.  According to
  Ref.\ \cite{Carenza:2019pxu}, the neutrino luminosity monotonically
  decreases after the SN explosion, while the axion luminosity shows
  an increasing behavior until $t_{\rm pb}\sim 5\ {\rm sec}$ and stays
  almost constant until $t_{\rm pb}\sim 10\ {\rm sec}$.  Thus, the
  signal-to-background ratio may be increased if we eliminate the data
  just after the SN explosion.  Because the detailed study of the time
  curves of the signal and backgrounds are beyond the scope of this
  article, we leave its study as a future work.}

\begin{figure}[t]
  \centering
  \includegraphics[width=0.7\textwidth]{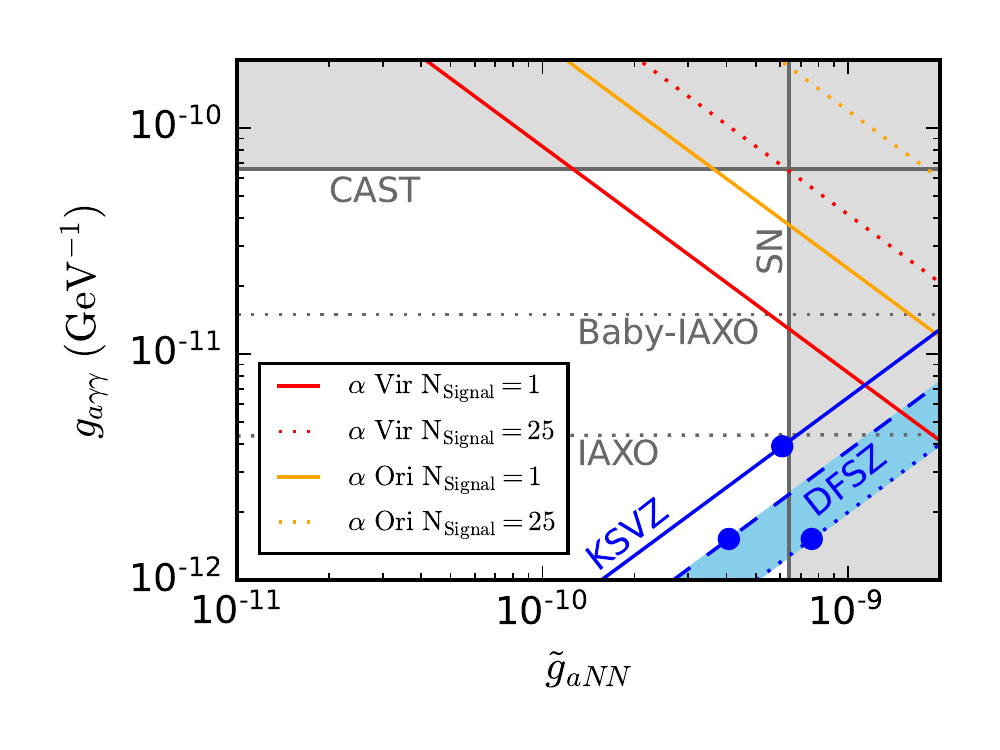}
  \caption{Contours of constant $N_{\rm signal}$ on $\tilde{g}_{ann}$
    vs.\ $g_{a\gamma\gamma}$ plane for Spica ($\alpha$ Vir, red) and
    Betelgeuse ($\alpha$ Ori, orange), for the ILD setup.  Here, we
    take $\Delta t_{\rm SN}=10\ {\rm sec}$, $\kappa_{\rm SN}=3$,
    $\theta=\frac{\pi}{2}$, and assume that $q^{-1}\gg\bar{L}$.  The
    vertical line shows the upper bound on $\tilde{g}_{aNN}$ from
    the SN cooling \cite{Carenza:2019pxu} (See Eq. \eqref{bound_gann}).  The solid and dotted horizontal lines indicate
    the CAST bound \cite{CAST:2017uph} and expected sensitivities of Baby-IAXO \cite{IAXO:2019mpb} and and
    IAXO \cite{IAXO:2019mpb}, respectively, assuming $m_a\lesssim 10\ {\rm meV}$.  The
    predictions of the QCD axion models in Eqs.\ \eqref{gaxx} --
    \eqref{gann_DFSZ} are shown in blue lines.  (The solid line is for
    the KSVZ model, and dashed and dotted lines are for DFSZ model
    with $\tan\beta=0$ and $\infty$, respectively.  The band in
    skyblue indicates the region predicted by the DFSZ model.  The
    blobs on the blue lines indicate the point at which $m_a=10\ {\rm
      meV}$.)}
  \label{fig:coupling_ILD}
\end{figure}

In Fig.\ \ref{fig:coupling_ILD}, we show the contour of constant
$N_{\rm signal}$.  On the same figure, we also show the parameter
region excluded by the SN cooling or by the CAST experiment.  We can
see that, if $d_{\rm SN}\lesssim 200\ {\rm pc}$ (roughly corresponding
to the distance to Betelgeuse), the number of signal can become $O(1)$
or larger in the parameter region which is still viable.  We can see
that, with the setup adopted in Fig.\ \ref{fig:coupling_ILD}, it is
difficult to reach the parameter region suggested by the QCD axion.
However, wide variety of ALPs may show up in the string theory and the
SN axion search of our proposal may be able to access some of those.
In addition, the SN axion search of our proposal may also reach a part
of parameter region covered by Baby-IAXO (and IAXO) experiment.  Thus,
if an ALP signal is found by Baby-IAXO, a joint analysis of the
results of Baby-IAXO and the SN axion search (if performed) will give
us deeper understanding of the ALP.

\begin{figure}[t]
  \centering
  \includegraphics[width=0.7\textwidth]{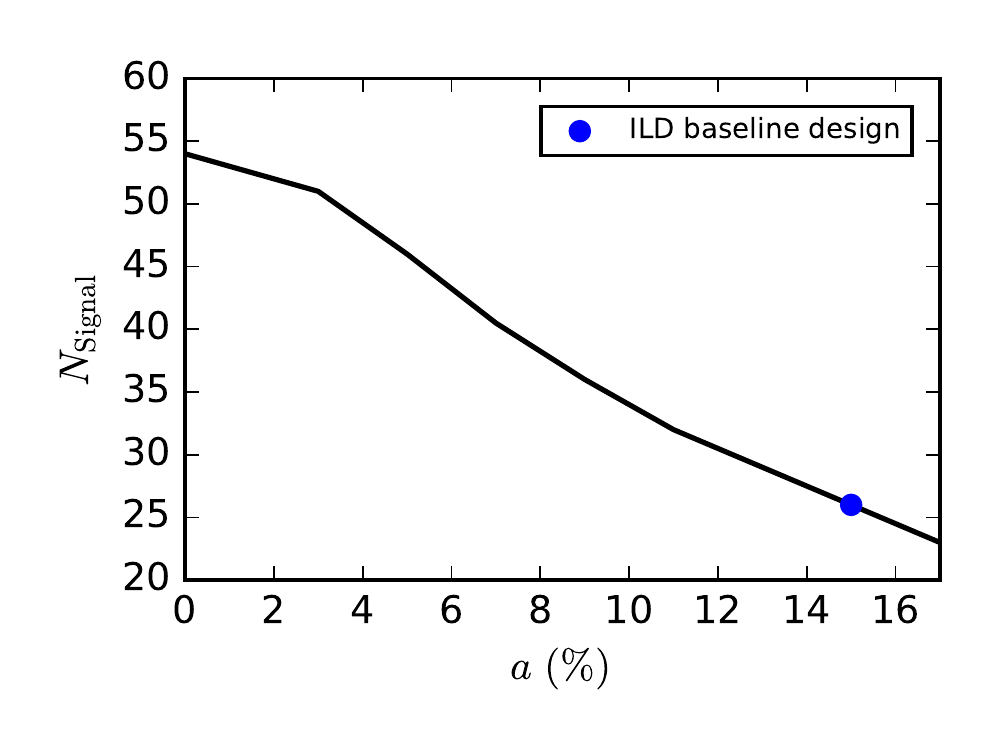}
  \caption{Number of signal as a function of the 
    detector-resolution parameter $a$ (see Eq.\ \eqref{dE(general)}) for the case of the ILD setup except for the ECAL resolution.
    Here, we take $\tilde{g}_{aNN}=6.4 \times 10^{-10}$,
    $g_{a\gamma\gamma}=6.6\times 10^{-11}\ {\rm GeV}^{-1}$, $d_{\rm
      SN}=77\ {\rm pc}$, $\Delta t_{\rm SN}=10\ {\rm sec}$,
    $\kappa_{\rm SN}=3$ and $\theta=\pi/2$.}
  \label{fig:resdep_ILD}
\end{figure}

As we have seen, the background spectrum is strongly affected by the
detector resolution. If a better resolution than the one given in
Eq.\ \eqref{de(ILD)} is realized, we may reduce the number of
background with lower $E_{\rm cut}$.  We assume the following form of
the detector resolution:
\begin{align}
  \frac{\delta E}{E} = \frac{a}{\sqrt{E_{\rm GeV}}},
  \label{dE(general)}
\end{align}
and see how $N_{\rm signal}$ depends on the detector resolution with
determining $E_{\rm cut}$ by solving Eq.\ \eqref{Ecut}.  The number of
signal increases with an ECAL with better resolution.  With the ECAL
based on ${\rm PbWO_4}$, which is used in the CMS, $a\sim 3\ \%$ can
be realized \cite{CMS:2013lxn}.  With such an ECAL, for example, the number of signal
can be increased by the factor of $\sim 2$ compared to the case with
the tungsten-based ECAL.

So far, we have adopted axion and neutrino spectra calculated with
assuming that the axion emission is subdominant for the SN cooling.
If the axion-nucleon-nucleon coupling becomes large, significant
amount of the energy is carried away by the axion and the cooling
process of the SN is affected. The axion and neutrino spectra for the
case of $L_a\sim L_\nu$ has been studied in
Ref.\ \cite{Fischer:2021jfm}; if $L_a\sim L_\nu$, the axion spectrum
may become softer than the one we adopt.  With such a softer spectrum,
the number of photon after imposing the cut may be reduced compared to
our previous estimation.  We have checked that, even if we use the
axion spectrum given in Ref.\ \cite{Fischer:2021jfm}, the number of
signal event with $E\gtrsim E_{\rm cut}$ can be as large as
a few for $d_{\rm SN}\sim 100\ {\rm pc}$.

%
%

One caveat for the case with the ILC is on the detector operation.  In
the ILC experiment, the bunches of $e^+$ and $e^-$ collide every $\sim
200\ {\rm millisec}$ and the bunch train is about $1\ {\rm millisec}$
long.  For $\sim 199$ out of $200\ {\rm millisec}$, the detector is
planned to be turned off in the normal operation (so-called
``power-pulsing'').  The SN axion search is hardly performed if the
livetime is reduced down to $\sim 0.5\ \%$ as in the case of the
normal operation.  Thus, a special detector operation dedicated for
the SN axion search is necessary.  In particular, the detector (at
least the ECAL) should be continuously turned on during the time
window of the SN.  In addition, the energy of the photon of our
interest is $\sim 100\ {\rm MeV}$; the energy threshold of the ECAL
should be low enough to detect such a photon.  With the continuous
operation of the detector, the power consumption may be an issue
\cite{Behnke:2013lya}.  For the detection of the SN axion, the vertex
and tracking detectors are irrelevant and they can be fully turned
off, which may help to reduce the power consumption during the SN
axion search.\footnote
{The event rate due to the SN neutrino is $\sim 10\ {\rm kHz}$ for an
  optimistic choice of parameters, which is of the same order of
  magnitude of the total collision rate in the normal operation of the
  ILC.  Thus, we expect that the data size of the whole event due to
  the nearby SN (including the background) is within the capacity.
  Most of the background can be removed via an off-line analysis by
  imposing the cut on the energy deposit.  }

\subsection{Detection at the LHC detectors}

Next, we comment on the LHC detectors, i.e., the ATLAS and the CMS.

As shown in Table \ref{table:detectors}, the ATLAS has weaker magnetic
field and shorter effective path length than the CMS.  The expected
number of the signal event is small and the discovery of the signal of
the axion emission is challenging at the ATLAS even with the most
optimistic choices of the axion couplings and $d_{\rm SN}$.  The CMS
has stronger magnetic field than the ATLAS.  A naive estimation of the
number of signal based on Eq.\ \eqref{N_gamma} gives $N_\gamma\sim
O(10)$ with the most optimistic choices of parameters.  For the
detection of the SN axion by the ATLAS or the CMS detector, if
performed, a new trigger dedicated for the SN axion signal is
necessary; the trigger should be replaced at the time of the pre-SN
alert.

However, for the cases of the LHC detectors, the total amount of the
material in the tracker region is sizable.  From the beam pipe to the
outside of the tracker region, the total material budget of the ATLAS
detector is $\sim 0.3-2$ radiation length for $|\eta|=0-4$ (with
$\eta$ being the pseudorapidity) in the high luminosity scenario of
the LHC (HL-LHC) \cite{ATL-ITK-PROC-2020-008}.  For the case of the
CMS, it is $\sim 0.3-1.6$ radiation length for $|\eta|=0-1.5$ and
$\sim 1.6-0.8$ radiation length for $|\eta|=1.5-3$
\cite{LaRosa:2021cof}. Thus, the photon may not be regarded as a free
particle in the central region and the conversion rate of the SN axion
to the photon is suppressed.

Another possibility to use the LHC detectors for the SN axion
detection is to look for the photon originating from the SN axion
using the $e^+e^-$ pair converted from the photon by the tracker
material.  Because of the relatively large material budget of the LHC
inner detectors, the photon from the SN axion is, if produced,
converted to the $e^+e^-$ pair with high probability, which may be
regarded as a signal of the SN axion.  The detailed calculation of the
event rate of such $e^+e^-$ process is beyond the scope of this
article and we leave it as future project.

\section{Conclusions and Discussion}
\label{sec:conclusions}
\setcounter{equation}{0}

We have discussed the possibility to observe the axion emission from a
nearby SN, which may occur in the future, using collider
detectors. The axion produced in association with the SN event can be
converted to the photon by the strong magnetic field in the central
region of the detector and the photon can be detected by the ECAL
surrounding the central region. We have calculated the number of
signal event in existing and proposed detectors in the LHC and the ILC
experiments. For the detection of the signal, a special collider
operation dedicated for the SN axion signal is necessary at the time
of a nearby SN which can be known in advance by the pre-SN alert:
\begin{enumerate}
\item At the time of the pre-SN alert, the beam should be stopped to
  make the detector environment quite.
\item The detector operation should be switched to the one dedicated
  for the SN axion search.
\item Then, we just have to wait for the SN axion to come.  If a
  sizable number of photons are observed during the time window of the
  SN, it is an evidence of the axion emission from the SN.
\end{enumerate}
We have seen that, with an optimistic choices of parameters, the ILC
detectors may be used to observe the axion from a nearby SN.  The SN
axion search of our proposal may access ALPs suggested by string
models, which are not excluded yet, while it is difficult to reach the
parameter region of the QCD axion.

Several comments are in order:
\begin{itemize}
\item In the SN axion search of our proposal, only the ``lower" half of
  the ECAL is used for the signal detection, and no signal is expected
  in the ``upper" half.  Thus, the study of the number of event in the
 ``upper" half of ECAL will provide a reliable estimation of the
  background.
\item So far, we have concentrated on the detection of the SN axion
  from a nearby SN.  For such a purpose, the neutrino-induced event
  are regarded as background.  However, the study of the
  neutrino-induced events may be also interesting. For example, with
  the collider detectors, we may obtain information about the energy
  spectrum of the neutrino (in particular, $\nu_e$).  We may also
  learn the time-dependence of the neutrino emissivity from the SN.
  Such information, if obtained, can be used to acquire deep insights
  into the physics of the SN.
\end{itemize}

Our proposal of the SN axion detection is low-cost assuming that the
change of the detector operation can be done at the software level.
There is (almost) no effect on the regular program of the collider
experiment.  Thus, we suggest each detector collaboration to prepare
in advance a detailed procedure for a nearby SN even though it rarely
happens.

\vspace{2mm}
\noindent{\it Acknowledgments:} 
The authors are grateful to 
Yutaro Iiyama,
Toshio Namba
and Taikan Suehara
for useful discussion and comments.
This work was supported
by JSPS KAKENHI Grant Nos.\ 
20H01911 (SA), 
16H06490 (TM), 18K03608 (TM), and 22H01215 (TM),
and also by the JSPS Fellowship No.\ 21J20445 (YK).


\bibliographystyle{jhep}
\bibliography{ref}


\end{document}